\documentclass{article}

\usepackage{spconf,amsmath,graphicx,acro, booktabs,tabularx,subcaption,amssymb,bm,svg}
\acsetup{single}

\DeclareAcronym{F18}{
	short = \textsuperscript{18}F,
	long=Fluorine 18,
}

\DeclareAcronym{Ga62}{
	short=\textsuperscript{62}Ga,
	long=gallium 62,
}

\DeclareAcronym{Rb82}{
	short=\textsuperscript{82}Rb,
	long=rubidium 82,
}

\DeclareAcronym{PET}{
	short=PET,
	long=positron emission tomography,
}

\DeclareAcronym{SPECT}{
	short=SPECT,
	long= Single Photon Emission Computed Tomography,
}

\DeclareAcronym{GAN}{
	short=GAN,
	long=generative adversarial network,
	long-plural-form = generative adversarial networks,
}

\DeclareAcronym{LS}{
	short=LS,
	long=least squares,
	long-plural-form = least squares,
}

\DeclareAcronym{GATE}{
	short=GATE,
	long=Geant4 Application for Tomography Emission,
}

\DeclareAcronym{MC}{
	short=MC,
	long=Monte Carlo,
}

\DeclareAcronym{PSF}{
	short=PSF,
	long=point spread function,
}

\DeclareAcronym{VAE}{
	short=VAE,
	long= variational autoencoder,
        long-plural-form = variational autoencoders,
}

\DeclareAcronym{LSTM}{
	short=LSTM,
	long=long short-term memory,
}

\DeclareAcronym{ViT}{
	short=ViT,
	long=vision transformer,
}

\DeclareAcronym{NN}{
	short=NN,
	long=neural network,
}

\DeclareAcronym{GNN}{
	short=GNN,
	long=graph neural network,
}

\DeclareAcronym{GRU}{
	short=GRU,
	long=gated recurrent unit,
}

\DeclareAcronym{GCN}{
	short=GCN,
	long=graph convolutional network,
}

\DeclareAcronym{ANN}{
	short=ANN,
	long= artificial neural network,
}

\DeclareAcronym{CNN}{
	short=CNN,
	long=convolutional neural network,
}

\DeclareAcronym{RNN}{
	short=RNN,
	long=recurrent neural network,
        long-plural-form = recurrent neural networks,
}

\DeclareAcronym{FCNN}{
	short=FCNN,
	long=fully-connected neural network,
}

\DeclareAcronym{MLP}{
	short=MLP,
	long=multilayer perceptron,
}	

\DeclareAcronym{LOR}{
	short=LOR,
	long=line of response,
	long-plural-form = lines of response,
}	

\DeclareAcronym{TOF}{
	short=TOF,
	long=time-of-flight
}

\DeclareAcronym{LM}{
	short=LM,
	long=list-mode
}

\DeclareAcronym{AI}{
	short=AI,
	long=artificial intelligence
}

\DeclareAcronym{MPNN}{
	short=MPNN,
	long=message passing neural network
}

\DeclareAcronym{IN}{
	short=IN,
	long=interaction network,
}

\DeclareAcronym{MIN}{
	short=MIN,
	long=modified interaction network,
}

\DeclareAcronym{MDN}{
	short=MDN,
	long=mixture density network,
}

\DeclareAcronym{2D}{
	short=2-D,
	long=two-dimensional,
}

\DeclareAcronym{3D}{
	short=3-D,
	long=three-dimensional,
}

\DeclareAcronym{1D}{
	short=1-D,
	long=one-dimensional,
}

\DeclareAcronym{PDF}{
	short=PDF,
	long=probability density function,
}

\DeclareAcronym{FWHM}{
	short=FWHM,
	long=full width at half maximum ,
}

\DeclareAcronym{RMSE}{
	short=RMSE,
	long=root-mean-square error,
}

\DeclareAcronym{MLEM}{
	short=MLEM,
	long=maximum-likelihood expectation-maximization algorithm,
}

\DeclareAcronym{SNR}{
	short=SNR,
	long=signal-to-noise ratio,
}

\DeclareAcronym{PSNR}{
	short=PSNR,
	long=peak signal-to-noise ratio,
}

\DeclareAcronym{MAE}{
	short=MAE,
	long=mean absolute error,
}

\DeclareAcronym{LHC}{
	short=LHC,
	long=large hadron collider,
}

\DeclareAcronym{XCAT}{
	short=XCAT,
	long=Extended Cardiac-Torso,
}

\DeclareAcronym{EM}{
	short=EM,
	long=expectation maximization,
}

\DeclareAcronym{castor}{
	short=CASToR,
	long=the Customizable and Advanced Software for Tomographic Reconstruction,
}

\DeclareAcronym{GT}{
	short=GT,
	long=ground truth,
}

\DeclareAcronym{ROI}{
	short=ROI,
	long=region of interest,
}

\DeclareAcronym{STD}{
	short=STD,
	long=standard deviation,
}

\DeclareAcronym{LXe}{
	short=LXe,
	long=liquid xenon,
}

\DeclareAcronym{phsp}{
	short=PhSp,
	long=Phase Space,
}

\usepackage{enumitem}
\setlist{nosep, leftmargin=14pt}

\usepackage{mwe} 

\usepackage{color}

\usepackage[
	style=ieee,
	dashed=false,
	maxbibnames=20,
	minbibnames=2,
	maxcitenames=1,
	mincitenames=1,
	backend=biber,
	defernumbers=false,
	sorting=none,
	useprefix=false
	]{biblatex}

\title{Fast-Track of F-18 Positron paths simulations using GANs}
\name{
	\begin{tabular}{@{}c@{}}
		Youness Mellak\textsuperscript{1}, Konstantinos Chatzipapas\textsuperscript{1}, Alexandre Bousse\textsuperscript{1}, Catherine Chez-Le Rest\textsuperscript{2},\\
		  Dimitris Visvikis\textsuperscript{1}, Julien Bert\textsuperscript{1}%
	\end{tabular}
}

\address{
    \textsuperscript{1}LaTIM, Inserm UMR 1101, Université de Bretagne Occidentale, Brest, France.\\
    \textsuperscript{2}Nuclear Medicine Dept, University of Poitiers, University Hospital of Poitiers, Poitiers, France
}
\begin{document}

%
\maketitle

\begin{abstract}

In recent years, the use of \ac{MC} simulations in the domain of Medical Physics has become a state-of-the-art technology that consumes lots of computational resources for the accurate prediction of particle interactions. The use of \ac{GAN} has been recently proposed as an alternative to improve the efficiency and extending the applications of computational tools in both medical imaging and therapeutic applications. This study introduces a new approach to simulate positron paths originating from \ac{F18} isotopes through the utilization of \acp{GAN}. The proposed methodology developed a pure conditional transformer \ac{LS}-\ac{GAN} model, designed to generate positron paths, and to track their interaction within the surrounding material. Conditioning factors include the predetermined number of interactions, and the initial momentum of the emitted positrons, as derived from the emission spectrum of \ac{F18}. By leveraging these conditions, the model aims to quickly and accurately  simulate electromagnetic interactions of positron paths. Results were compared to the outcome produced with \ac{GATE}~\ac{MC} simulations toolkit. Less than 10 \% of difference was observed in the calculation of the mean and maximum length of the path and the 1-D \ac{PSF} for three different materials (Water, Bone, Lung).
\footnote{The code corresponding to this study can be found at: \url{https://github.com/Mellak/Particles_Tracking_GAN}.}

\end{abstract}


\section{Introduction}

Simulations and modeling are essential tools in medical physics clinical and pre-clinical applications, both for therapeutic and imaging purpose. \Ac{MC} simulations enable the optimization of image acquisition parameters, image calibration, estimation of dose distribution, and detection of system enhancement. Due to the stochastic nature of this method, results with sufficient statistical uncertainty require significant calculation time and computing power. This limits its use in clinical routine but also in research. For almost 50 years, researchers have been working on improving the efficiency of \Ac{MC} simulations. Particle tracking is particularly time-consuming in simulation because billions of particles must be tracked through matter. Recently, a deep learning approach have been proposed to learn the particle distribution escaping the patient in \ac{SPECT} MC simulation \cite{sarrut2021SPECT}. This proof of concept, based on \acp{GAN}, has been avoided particle tracking through the patient and then improve the computation time. Similarly, in \cite{sarrut2023annihilation}, a \acp{GAN} model was used to surrogate the annihilation of the positron by directly producing the resulted pairs of gammas. This model, that include temporal information, was able to save time for any medical imaging simulation using positron emitter. Even though this method generates accurate statistics of back-to-back photons, including scattering and absorption, it does not track the full length of the path for each particle. The simulation of the full particle's path as it interacts with matter incorporates several difficulties. In the case of positrons, the number of possible types of interactions is relatively high, (Bhabha scattering, ionization, Bremsstrahlung effects, Cherenkov radiation, and positron-electron annihilation). Additionally, positron's kinetic energy depends on its emitting isotope. As a result, the path length and number of interactions of each positron track depends on its energy. The use of a fully-connected generator/discriminator produced promising results in generating the state of particles at any specific point given the starting position \cite{sarrut2023annihilation}. However, this model lacks information regarding the path followed, such as the continuous energy decrease of the particle along its path, and the position where the positron annihilates. In this work, we propose a novel approach to address this limitation by treating the task of path generation as a time series problem.

Several studies incorporate the \ac{GAN} framework into a temporal context. The initial approach (C-RNN-GAN) \cite{mogren2016c} applied a \ac{GAN} architecture directly to the sequential data, utilizing \ac{LSTM} networks for both the generator and  the discriminator. Recurrent Conditional \ac{GAN} (RCGAN) \cite{esteban2017real} follows a similar strategy with limited architectural adjustments. A recent work suggested conditioning on  the time stamp information to handle irregular sampling. These approaches rely solely on binary adversarial feedback, which could potentially lack to capture temporal dynamics in training data. An alternative architecture namely TimeGAN, was introduced \cite{yoon2019time}. It is explicitly trained to preserve temporal dynamics through a step-wise supervised loss and an embedding network, demonstrating effectiveness in capturing temporal relationships and accommodating mixed-data scenarios. However, traditional \acp{RNN} are difficult to train and may suffer from vanishing gradients. In contrast, transformers offer a solution to these drawbacks and have shown impressive performance, particularly in handling long-range series. 

Our work is based on a pure-transformer \ac{LS}-\ac{GAN} architecture, described in \cite{li2022tts}, capable of generating high quality time series. We focus on simulating the trajectories and interactions of positrons within various homogeneous materials. Each positron trajectory is characterized by its energy and the number of steps taken, serving as input conditions for the model. Furthermore, this study serves as a proof-of-concept, including a complete validation of the \ac{GAN} model, as compared with state-of-the-art \ac{MC} simulations of positrons emitted by \ac{F18}.

\section{Materials and Methods}

\subsection{Proposed Architecture}

A positron path with a maximum of $N$ interactions (including the emission) takes the form of an  array $\mathbf{p} \in \mathbb{R}^{N\times 4}$ where for all interaction index $n=0,\dots, N-1$, $[\mathbf{p}]_n = [e_n,\mathbf{r}_n]$ where $\mathbf{r}_n  = [x_n,y_n,z_n] $ and $e_n$ are respectively the 3-D location at the $n$th interaction and the positron's energy after the $n$th interaction, $[e_0,\mathbf{r}_0]$ being the initial state of the positron. We assume that $\mathbf{r}_0=[0,0,0]$. The sequence $(e_n)$ is nonnegative and  decreasing as each interaction results in a loss of energy, and such that the initial energy $e_0>0$ is predetermined.
The generator is a mapping \(G_{\bm{\theta}} = [G_{\bm{\theta}}^\text{en}, G_{\bm{\theta}}^\text{geom}]\colon \mathcal{Z} \times \mathcal{Y} \to \mathcal{C}\), where $\bm{\theta}$ is the parameters of the generator, \(\mathcal{Z}\) is the latent space with a latent variable \(\mathbf{z}\) drawn from the probability density function \(p(\mathbf{z})\), \(\mathcal{Y}\) is the space of condition vectors, \(\mathcal{C}\) is the space of the data, \(G_{\bm{\theta}}^\text{geom}(\mathbf{z}, \mathbf{y}) \in \mathbb{R}^{N \times 4}\) represents the geometric path, and \(G_{\bm{\theta}}^\text{en}(\mathbf{z}, \mathbf{y}) \in \mathbb{R}^N\) represents the corresponding energy sequence.The generator is thus conditioned on both the latent variable \(\mathbf{z}\) and the condition vector \(\mathbf{y}\). The notation \(G_{\bm{\theta}}^\text{geom}(\mathbf{z}, \mathbf{y})\) and \(G_{\bm{\theta}}^\text{en}(\mathbf{z}, \mathbf{y})\) reflects the dependence on both \(\mathbf{z}\) and \(\mathbf{y}\).

\begin{figure}[htb]
  \centering
  \includegraphics[width=8.5cm]{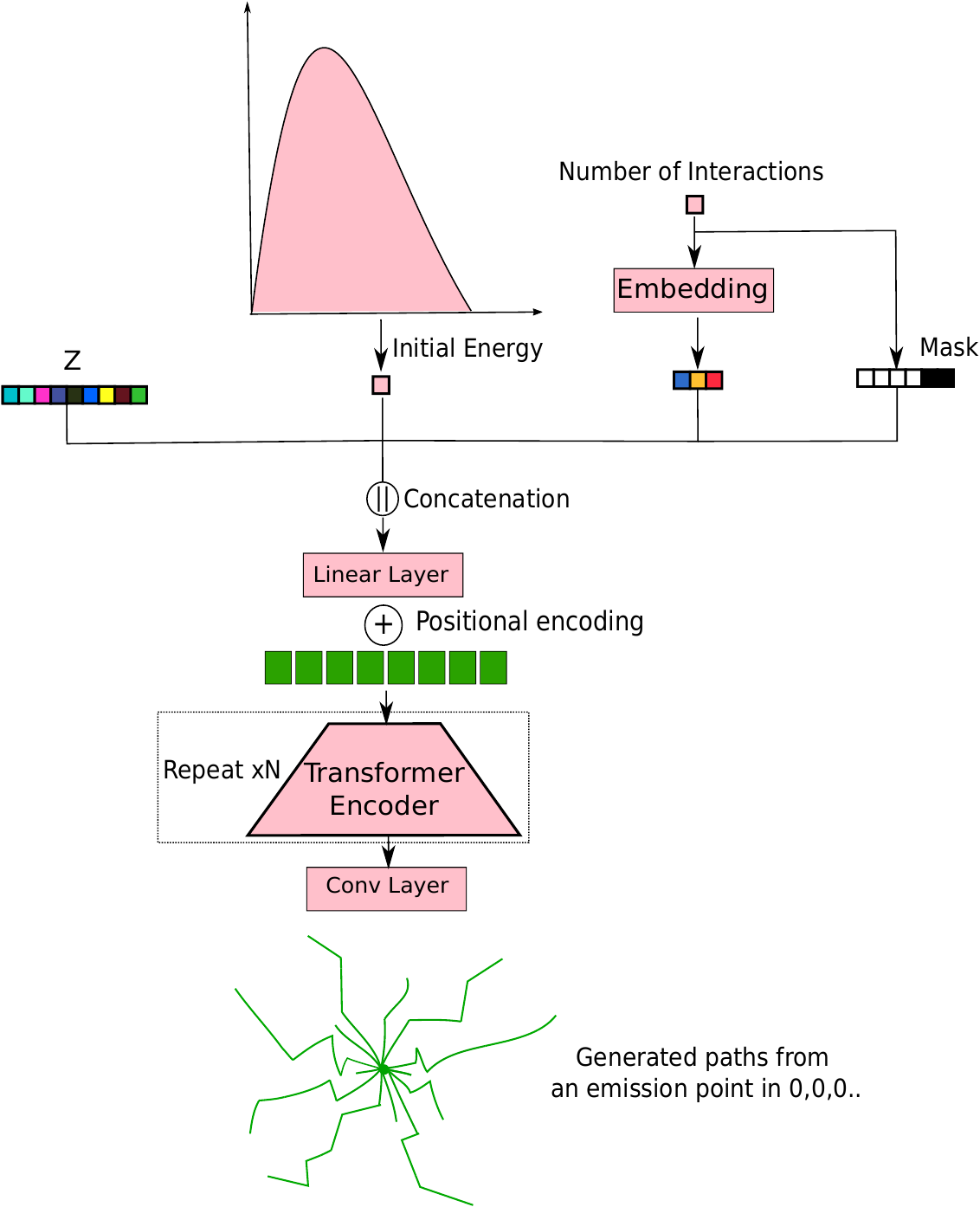}
  \caption{The generator architecture begins by taking the concatenated input of various components: 1)the embedded number of interactions, 2) a mask indicating 0 values for features beyond the specified number of interactions, 3) the initial energy, and 4) a randomly sampled vector from N(0,1). This vector undergoes mapping and reshaping to conform to the sequence shape. Subsequently, a transformer encoder generates new embedding for each sequence point, followed by a convolutional layer that transforms features from the embedding dimension to the dimensions of real data features. }
  \label{fig:generator}
\end{figure}

The generator in our architecture is based on a Transformer encoder that takes as input vectors of specified dimensions that include concatenate random vector and embedded conditions and produces a sequence with the same shape as the input vectors \cite{vaswani2017attention}. Afterward, this sequence is transformed to match the real data dimension using 1-D convolutions see Figure \ref{fig:generator}.

\begin{figure}[htb]
  \centering
  \includegraphics[width=8.5cm]{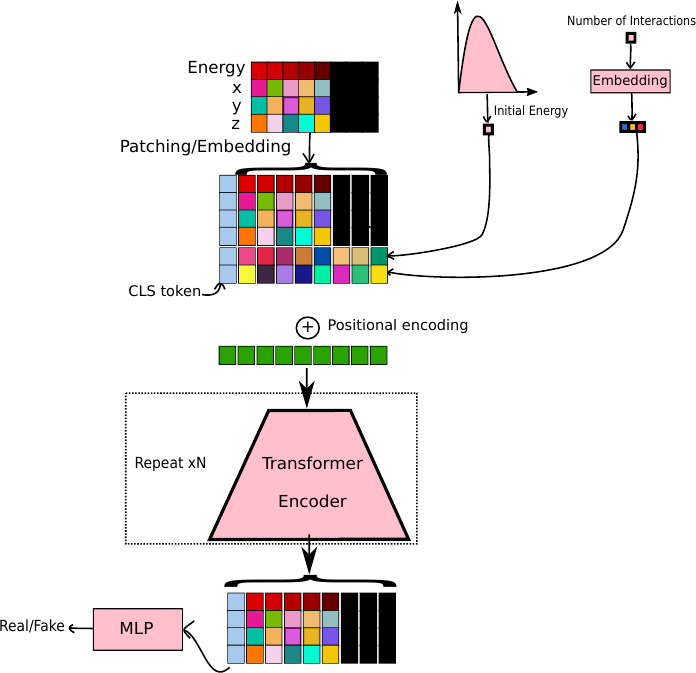}
  \caption{The discriminator operates by taking the input of the embedded initial energy and the number of interactions along the path to be classified. This input is then processed through a model akin to \ac{ViT}, where each patch represents a distinct point and encompasses features such as Energy, x, y, and z. The discriminator's objective is to classify the input as either real or fake.}
  \label{fig:discriminator}
\end{figure}

The discriminator is inspired from \ac{ViT} \cite{dosovitskiy2020image} and it is a binary classifier that is used to distinguish between real and synthetic data. It is a mapping \(D_{\bm{\phi}} \colon \mathcal{C} \to [0,1]\), where \(D_{\bm{\phi}}\) assigns each element in \(\mathcal{C}\) (positron paths) to a scalar value between 0 and 1 and $\bm{\phi}$ represent parameters of the discriminator. It takes as input a path concatenated with embedded conditions and outputs a single scalar value that represents the probability of the input being real or synthetic see Figure \ref{fig:discriminator}. 

The original TTS-GAN model \cite{li2022tts} was trained using the \ac{LS}-\ac{GAN} loss function \cite{mao2017least}, which is a modification of the original \ac{GAN} loss function. The generator and discriminator are updated alternately using back-propagation. The generator loss is defined as:
\begin{equation}\label{eq:lsgan_g_conditioned}
L_{\text{g}}(\bm{\phi}, \bm{\theta}, \mathbf{y}) = \frac{1}{2} \mathbb{E}_{\mathbf{z} \sim p(\mathbf{z})}\left[\left(D_{\bm{\phi}}(G_{\bm{\theta}}(\mathbf{z}, \mathbf{y})) - 1\right)^2\right]
\end{equation}
and the discriminator loss defined as:
\begin{align}
L_{\text{d}}(\bm{\phi}, \bm{\theta}, \mathbf{y}) = {} & \frac{1}{2} \mathbb{E}_{\mathbf{x} \sim p_{\text{data}}(\mathbf{x})}\left[(D_{\bm{\phi}}(\mathbf{x}, \mathbf{y}) - 1)^2\right] \nonumber \\ 
& + \frac{1}{2} \mathbb{E}_{\mathbf{z} \sim p(\mathbf{z})}\left[(D_{\bm{\phi}}(G_{\bm{\theta}}(\mathbf{z}, \mathbf{y})))^2\right]
\label{eq:lsgan_d_conditioned}
\end{align}
The TTS-GAN model is trained by alternating between minimization of $L_{\text{g}}$ with respect to $\mathbf{\theta}$  and minimization of $L_{\text{d}}$ respect to $\mathbf{\phi}$ until the generator is able to generate synthetic data that is indistinguishable from the real data by the discriminator.

The positron paths differ in length and number of interactions, ranging from $N=3$ to 18. As our model is a one-shot generator, it should be capable of generating paths with different numbers of interactions. To do so, we help the model by passing a mask as input. The mask makes features out of the number of interactions set to zeros, before passing them to the discriminator.

In our context we can help the model to learn the characteristics faster by introducing two terms in the loss function: The generated energy of the particle should be positive and decrease as it advances, and in the last interaction it should be 0. The regularized loss function is given by:
\begin{equation}
\begin{split}
    &\text{DecrLoss}(\bm{\theta}) = \\
    &\mathbb{E}_{\mathbf{z} \sim p(\mathbf{z})} \left[ \sum_{j=2}^{N} \text{ReLU}\left([G_{\bm{\theta}}^\text{en}(\mathbf{z},\mathbf{y})]_{j-1} - [G_{\bm{\theta}}^\text{en}(\mathbf{z},\mathbf{y})]_{j}\right) \right]
\end{split}
\end{equation}

\begin{align}
    \text{NegLoss}(\bm{\theta}) &= \mathbb{E}_{\mathbf{z} \sim p(\mathbf{z})} \left[ \sum_{j=1}^{N} \text{ReLU}\left(-[G_{\bm{\theta}}^\text{en}(\mathbf{z},\mathbf{y})]_{j}\right) \right]
\end{align}
where $\text{ReLU}(t)$ denotes the Rectified Linear Unit function, defined as $\text{ReLU}(t) = \max(0, t)$, and $[G_{\bm{\theta}}^\text{en}(\mathbf{z},\mathbf{y})]_{j} \in \mathbb{R}$ represent the energy of j-th point of the path.

To ensure the generator produces a particle with a desired initial direction, we add the cosine similarity between the normalized initial vector and the provided condition for the desired direction. A low cosine value signals alignment with the intended direction. Including this cosine loss in training helps the model create sequences with reliable directional information, enabling control over the generated direction. The cosine similarity between the initial vectors, represented by the first two points of the path \(\left[G_{\bm{\theta}}^\text{geom}(\mathbf{z}, \mathbf{y})\right]_1 \in \mathbb{R}^3\) and \(\left[G_{\bm{\theta}}^\text{geom}(\mathbf{z}, \mathbf{y})\right]_2 \in \mathbb{R}^3\), and the direction condition \(\mathbf{v} \in \mathbb{R}^3\), an element of \(\mathbf{y}\), is expressed as:
\begin{equation}
    \text{SimLoss}(\bm{\theta}) = \frac{[G_{\bm{\theta}}^\text{geom}(\mathbf{z}, \mathbf{y})]_2 - [G_{\bm{\theta}}^\text{geom}(\mathbf{z}, \mathbf{y})]_1 \cdot \mathbf{v}}{\|[G_{\bm{\theta}}^\text{geom}(\mathbf{z}, \mathbf{y})]_2 - [G_{\bm{\theta}}^\text{geom}(\mathbf{z}, \mathbf{y})]_1\| \cdot \|\mathbf{v}\|}
\end{equation}
Where \(\cdot\) indicates the dot product, and \(\|\|\) signifies the Euclidean norm.
The final generator loss is defined as: 
\begin{multline}
    L_{fg}(\bm{\phi}, \bm{\theta}, \mathbf{y}) = L_{g}(\bm{\phi}, \bm{\theta}, \mathbf{y}) + \text{SimLoss}(\bm{\theta})  \\ + 0.005 \cdot \left(\text{DecrLoss}(\bm{\theta}) + \text{NegLoss}(\bm{\theta})\right)
\end{multline}

\subsection{Data}
Training data were produced using \ac{GATE} \ac{MC} \cite{strulab2003gate} simulations to model a point source located at (0,0,0) with the positron energy spectrum of \ac{F18}. This source is placed within an analytical phantom that contains a sphere with a 5-cm radius. A \ac{phsp} actor records the path followed by positrons step-by-step. We conduct simulations in three different materials---water, lung, and bone---resulting in approximately 10,000 events for each material. The collected data was stored in a three dimentional numpy array storing: the number of events, the maximum number of interactions, and the features shape. Positrons produced by \ac{F18} can undergo a maximum of 18 interactions. The features in the dataset include the kinitic energy along with the coordinates ($x$, $y$, $z$) at any given specific point. In each epoch we augment data by rotating the full path around the origin of the emission, as well as by rotating the path around the axis defined by the first and the last interaction.
Three models were trained separately, each of each different material.

\section{Results}
\begin{figure}[htb]

\begin{subfigure}[b]{\linewidth}
  \centering
  \includegraphics[width=8.5cm]{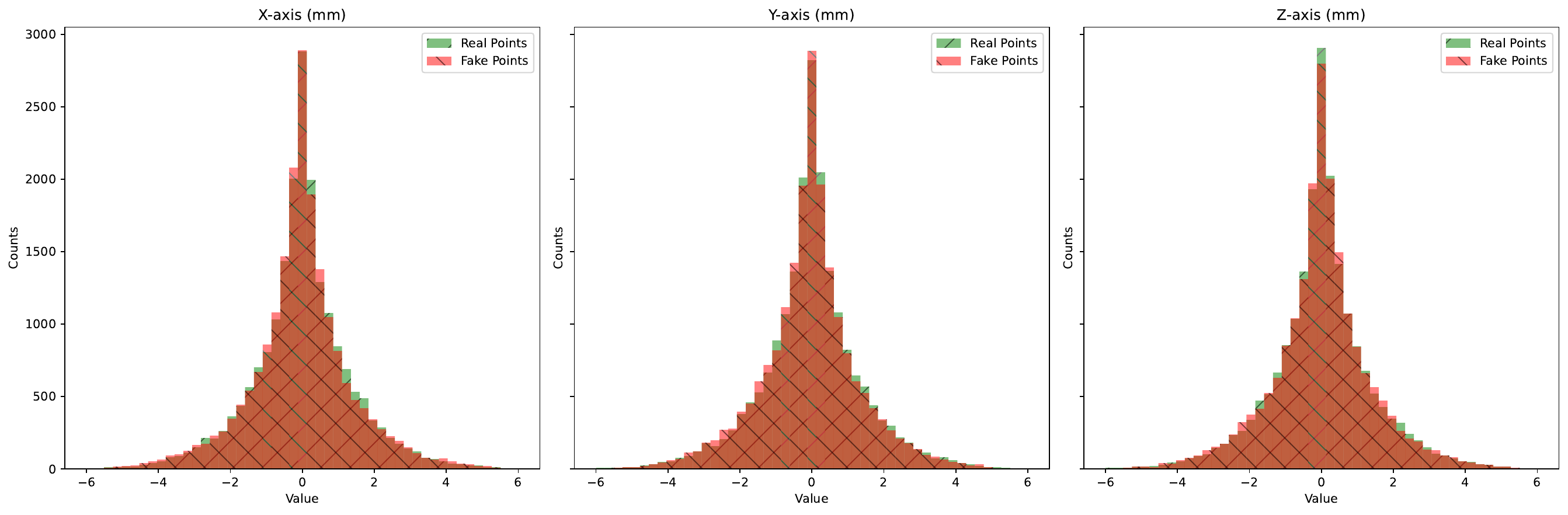}
  \caption{\Acp{PSF} in lung through different axes ($x$, $y$, and $z$).}
\end{subfigure}

\begin{subfigure}[b]{\linewidth}
  \centering
  \includegraphics[width=8.5cm]{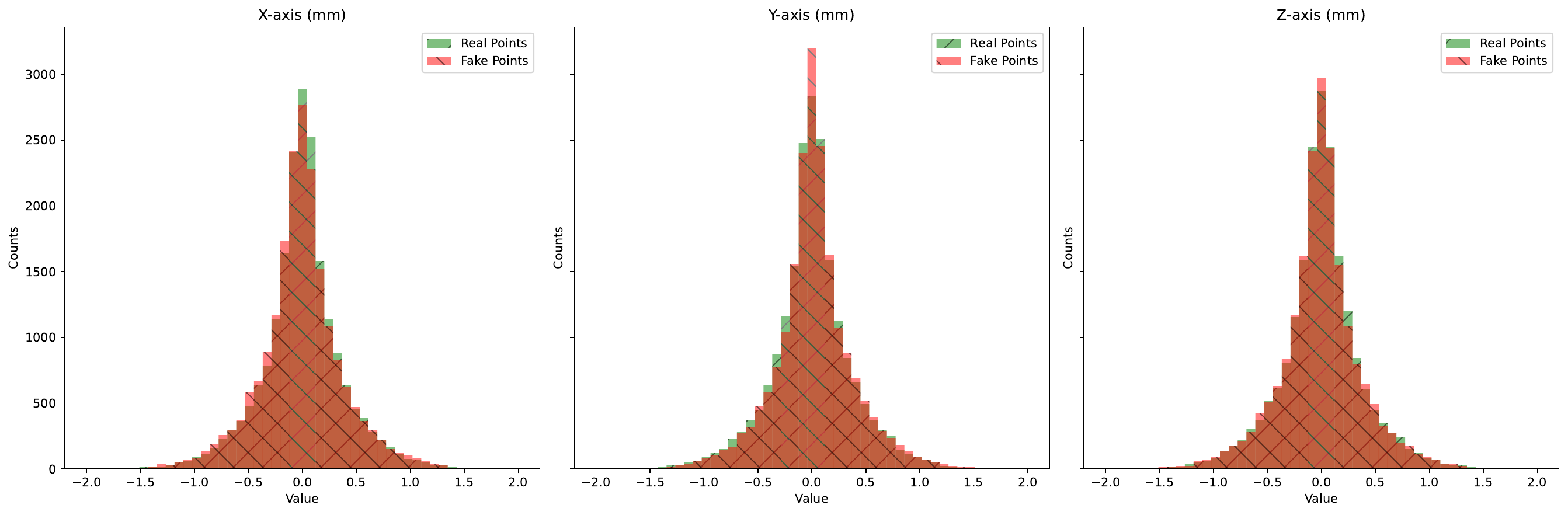}
  \caption{\Acp{PSF} in water through different axes ($x$, $y$, and $z$).}
\end{subfigure}

\begin{subfigure}[b]{\linewidth}
  \centering
  \includegraphics[width=8.5cm]{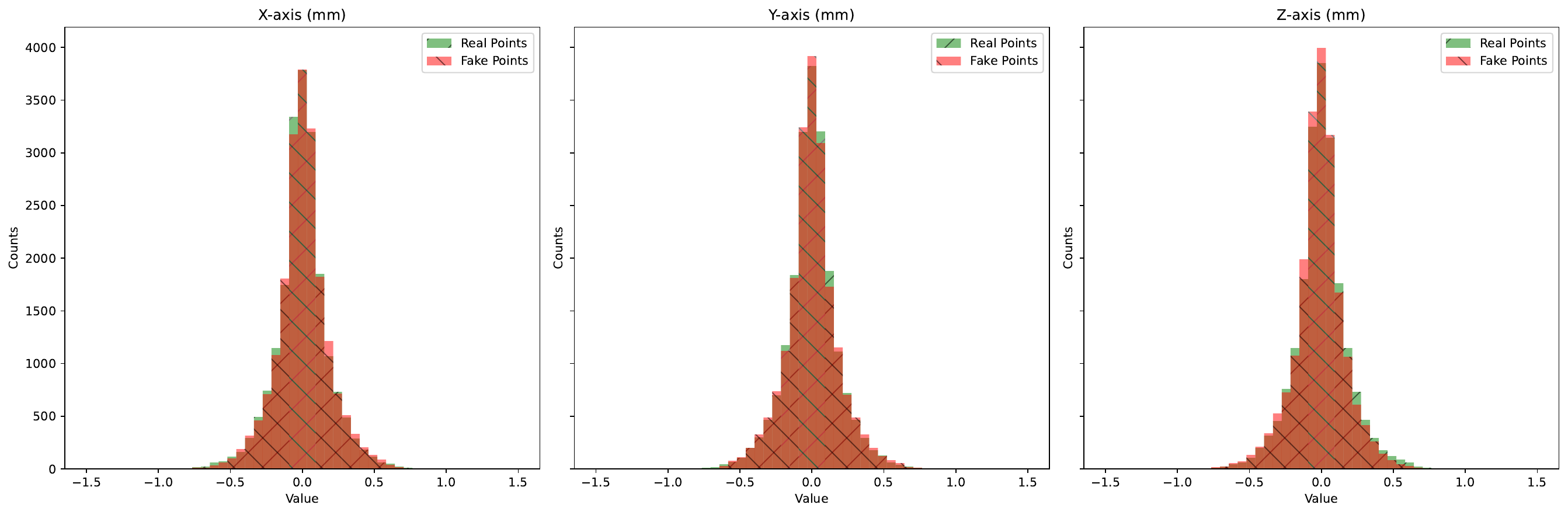}
  \caption{\Acp{PSF} in bone through different axes ($x$, $y$, and $z$).}
\end{subfigure}

\caption{Comparison of \ac{GATE} and \ac{GAN} \ac{PSF} in different materials through various axes.}
\label{fig:psf_results}

\end{figure}

Figure \ref{fig:psf_results} showcases the 1-D \ac{PSF} representation, which is generated by capturing the distribution of terminal points from paths produced by GATE (depicted in green) and this study's \ac{GAN} (depicted in red). These paths originate from the origin (0,0,0) along distinct $x$, $y$, and $z$ axes and through three different materials. Furthermore, a comprehensive comparative analysis of the mean and maximum radii of positron paths, is shown in detail in Table \ref{tab:mean_max_paths}. The \ac{GAN} model demonstrates proficiency in replicating paths with mean lengths ($R_\text{mean}$) similar to those generated by GATE. Difference less than 10\% is observed for the calculated ($R_\text{max}$).
\begin{table}[h]
    \centering
    \begin{tabularx}{\linewidth}{l*{4}{>{\centering\arraybackslash}X}}
        
        \cmidrule{3-4}
        & & $R_\text{mean}$ & $R_\text{max}$ \\
        \midrule
        \textbf{Water} & GATE & 0.52 mm & 2.13 mm \\
        & \ac{GAN} & 0.52 mm & 2.02 mm \\
        \midrule
        \textbf{Bone} & GATE & 0.25 mm & 1.07 mm \\
        & \ac{GAN} & 0.26 mm & 0.94 mm \\
        \midrule
        \textbf{Lung} & GATE & 1.92 mm & 7.82 mm \\
        & \ac{GAN} & 1.93 mm & 7.60 mm \\
        \bottomrule
    \end{tabularx}

    \caption{Mean and max path lengths obtained with GATE and GAN.}
    \label{tab:mean_max_paths}
\end{table}

The time it takes to perform a standard \ac{MC} simulation can differ based on various factors in its setup. Key parameters, such as tracking cuts, physics list, and the type of volume (whether it's voxelized or analytical), play a significant role in determining the computation time. In this study's experiments, we utilized GATE to simulate scenarios with three point sources of 0.2 MBq, placed in three different 5 cm-radius spheres corresponding to the three materials. During these simulations, GATE required $\approx$45 seconds to produce the desired number of paths. However, with the proposed method, it only takes 6 seconds, and this is achieved by using a batch size of 20,000 events. Since the proposed model is small (less than 1MB), it allows the generation of large batches (up to 400,000 events per batch).

\section{Discussion and conclusion}
In this study, a new approach was proposed for particle tracking within different type of materials, modeling each particle trajectory as a sequence. By conditioning the proposed \ac{GAN} model on the energy and the number of interactions, this method enables the generation of particle paths with a varying number of interactions. Remarkably, these generated paths closely approximate the original \ac{PSF} while achieving high computational efficiency. The utilization of a transformer architecture, coupled with carefully crafted loss function terms, proved instrumental to generate paths that exhibit a gradual decrease in energy, ultimately leading to the positron's annihilation. Furthermore, the inclusion of a cosine term in the loss function enhances our ability to control the initial direction of particles, similarly to the simulation of particle beams.

The model exhibits the capability to generate a large number of events with different lengths in one-shot manner. When dealing with a source different from the origin, efficient simulations are achieved by starting from the origin (0,0,0) and applying space shifts based on the emission map. For voxelized phantoms, a more time-consuming process is involved, as explained in \cite{sarrut2019generative}. However, our approach remains consistent; we conduct simulations in continuous space and subsequently discreet them to align with a voxel grid.

It's important to note that the current study focuses on generating paths for \ac{F18} positrons within homogeneous materials. Looking ahead, our future work aims to extend the model to simulate particle paths in different volumes. Additionally, we plan to explore the application of our model to other radionuclides such as \ac{Ga62} and \ac{Rb82}, which exhibit more significant positron ranges and  large numbers of interactions. This broader scope will enhance the versatility and applicability of our proposed methodology.

\section{Acknowledgments}
This work was performed within the framework of the MOCAMED project (ANR-20-CE45-0025), the SIRIC LYriCAN Grant INCa-INSERM-DGOS-12563, the LABEX PRIMES (ANR-11-LABX-0063), the LABEX CAMINLABS (ANR-10-LABX-07-01) within the program “Investissements d’Avenir” (ANR-11-IDEX-0007) operated by the ANR, the SECURE project (HORIZON-EURATOM-2021-101061230), and the POPEYE ERA PerMed 2019 project (ANR-19-PERM-0007-04). Views and opinions expressed are however those of the author(s) only and do not necessarily reflect those of the European Union or EURATOM. Neither the European Union nor the granting authority can be held responsible for them.

\printbibliography
\end{document}